\documentclass[11pt]{article}
\newcommand{\sss}{\vspace{.2in}}

\newcommand{\be}{\begin{equation}}
\newcommand{\ee}{\end{equation}}
\newcommand{\bea}{\begin{eqnarray}}
\newcommand{\eea}{\end{eqnarray}}
\usepackage[dvips]{graphicx}
\usepackage{dcolumn}
\usepackage{amsmath}
\usepackage{amssymb}
\usepackage{latexsym}
\usepackage{subfigure}
\usepackage{color}

\newcommand{\hp} {\hat p}
\newcommand{\hx} {\hat x}

\newcommand{\bean}{\begin{eqnarray*}}
\newcommand{\eean}{\end{eqnarray*}}

\newcommand{\bq}{\begin{equation}}
\newcommand{\ba}{\begin{eqnarray}}
\newcommand{\eq}{\end{equation}}
\newcommand{\ea}{\end{eqnarray}}

\begin{document}
~\hfill{\footnotesize IP/BBSR/01-13,~~\today}
\sss
\sss
\begin{center}
{\Large {\Large \bf Classical Limit of Time-Dependent Quantum Field Theory-a}}
{\Large {\Large \bf Schwinger-Dyson Approach}}
\end{center}
\vspace{.5in}
\begin{center}
{\large{\bf
\mbox{Fred Cooper}$^{a,}$\footnote{fcooper@lanl.gov}
\mbox{Avinash Khare}$^{b,}$\footnote{khare@iopb.res.in} and
   \mbox{Harvey Rose}$^{a,}$\footnote{har@lanl.gov}
 }}
\end{center}
\begin{center}
\vspace{.6in}
\noindent
a) \hspace*{.2in}
Theoretical Division, Los Alamos National Laboratory, Los Alamos, NM 87545\\
b) \hspace*{.2in}
Institute of Physics, Sachivalaya Marg, Bhubaneswar 751005, Orissa, India\\
\end{center}
\sss
\sss

\date{\today}
\begin{abstract}
We rewrite the Martin-Siggia-Rose (MSR) formalism  for the statistical
dynamics of  classical fields in a covariant second order form
appropriate for the statistical dynamics of relativistic field theory.
This second order formalism is related to a rotation of Schwinger's
closed time path
(CTP) formalism for quantum dynamics, with the main difference being that
certain vertices are absent in the classical theory.  These vertices
are higher order  in an  $\hbar$ expansion.  The structure
of the second order formulation of the Schwinger Dyson (S-D)  equations is
identical to that of the rotated CTP formalism  apart from initial conditions
on the Green's functions and the absence of these vertices.  We then
discuss self-consistent truncation schemes based on keeping certain
 graphs in the two-particle irreducible effective action
made up of bare vertices and exact Green's functions. 
\end{abstract}
%
\newpage
%
%

{\noindent\bf I. INTRODUCTION}

Recently there has been a lot of interest in connecting hot
relativistic
quantum field theory which is important for cosmology and heavy ion physics
 with classical field theory. Although there have been several
insightful studies
in perturbation theory \cite{ref:aarts1}, a systematic analysis has
not been given for the Schwinger Dyson equations.  One exception is
the work of Wetterich \cite{ref:EQT}, but that work only concerns itself with equal 
time correlation functions.  Since formalisms already exist for
obtaining Schwinger Dyson (S-D)  equations for both the quantum
\cite{ref:CTP} and classical problems \cite{ref:MSR}, we thought it
timely to review the MSR formalism, and cast it in a form
so that it could be directly compared with the now more familiar CTP
formalism \cite{ref:pli}. We show here that the second order formulation of
the S-D Equations  in the MSR formalism is
identical to that of the quantum theory (CTP formalism)  apart from initial conditions
on the Green's functions and the absence of those vertices which are
higher order in $\hbar$. 

\sss
 {\noindent\bf II. REVIEW OF FIRST ORDER IN TIME MSR FORMALISM}

In the paper of Martin-Siggia-Rose (MSR) \cite{ref:MSR},  an operator
formalism was developed which allows one to find the generating functional for
both the correlation and response functions for first order classical field equations
of the type:
\begin{equation}
\dot x(r,t) = A[x(r,t)]~,
\end{equation}
where $A[x(r,t)]$ is a local polynomial in the classical field $x(r,t)$.
In the work of MSR, $A[x(r,t)]$ could contain dissipative terms as well as 
prescribed noise terms.  The formalism presented in MSR is first order
in time derivatives and not apparently covariant. What we will show
is that there is a covariant subset of the MSR equations in
terms of which all the MSR Green's functions can be recovered. This
subset
can be derived from 
a second order Lagrangian formulation which can  be related to the $\hbar \rightarrow 0$ limit of the
CTP formalism of Schwinger and Keldysh. 

 For the statistical  classical field evolutions of $N$ interacting  
classical fields $\phi_a$, $a=1,2 \ldots N$   then 
$x(r,t)$ is the $2N$  component field consisting of $\phi_a$ and the canonical
momentum $\pi_a= {\dot \phi}_a$.  
\begin{equation}
  x = \left( \begin{array}{c} 
 \phi_a \\ \pi_a 
\end{array} 
\right) .
\end{equation}
If for example  we restrict ourselves to cubic interactions, then the
vector
$A$ is of the form
\begin{equation}
A_i = c_i (r,t) + d_{ij} x_j(r,t) + \frac{1}{2} e_{ijk} x_j(r,t) x_k(r,t)~,
\end{equation}
where $i= 1 \ldots 2 N $.  In the MSR formalism one  introduces the operator
\[{\hat x (r,t)} \equiv -{\delta \over \delta x(r,t)}~, \]
such that the commutation rule 
\begin{equation}
[x(r,t), {\hat x}(r',t)] = \delta (r-r') \label{eq:comm1}~,
\end{equation}
is true. Defining an operator Hamiltonian $H$
\begin{equation}
H(t) = \int~ dr' ~ {\hat x}_i(r',t)  A_i(r',t)~,
\end{equation}
the equations of motion can be written in the compact form
\begin{equation}
{\dot x} (r,t) = [x(r,t), H(t)]~.
\end{equation}
For  Eq. (\ref{eq:comm1}) to be true at all times one needs that ${\hat x}$
satisfies
\begin{equation}
 {d {\hat x} (r,t) \over dt} = [{\hat x}(r,t), H(t)]~,
\end{equation}
for consistency.  Therefore $\hat {x} (r,t) $ is a functional of $x(r,0)$ and
$\hat {x} (r,0) $. The formal solution to
these equations is given by (in what follows we suppress the spatial coordinate $r$) 
\begin{eqnarray}
x(t) && = U^{-1}(t,0) x(0) U(t,0) \nonumber \\
\hat {x} (t) && =  U^{-1}(t,0) {\hat x}(0) U(t,0)~, 
\end{eqnarray}
where
\[ U^{-1} (t,t_0) = T \exp[-\int_{t_0}^t H(t') dt']~; ~~  U (t,t_0) = T^\ast
\exp[\int_{t_0}^t H(t') dt'] ~. \]
 The meaning of the expectation value 
$\langle  x(t) {\hat x} (t')
\rangle $ is as follows. Given an initial probability function $P[x(0)]$ then
\begin{equation}
\langle  x(t) {\hat x} (t')
\rangle = \int dx(0)~ x[t,x(0)] ~~{\hat x} [t',x(0), {\hat x}(0)] ~P[x(0)]~.
\end{equation}
Wherever ${\hat x}(0)$ appears, it is replaced by 
\[ {\hat x}(0) \rightarrow - {\delta \over  \delta x(0)}~, \]
 and
it acts on everything to the right of itself. This definition of the extended
averaging procedure has three important properties.(1)
The average of a product of $x$'s agrees with the conventional definition.
(2)The time dependence of $\langle  x(t) {\hat x} (t')
\rangle $ is consistent with the above equations of motion.
(3)
The expectation value of a product of $x$ and ${\hat x}$ which has an ${\hat x}$
to the left vanishes if $P[x(0)]$ goes to zero fast enough at large $|x|$.

The last property is crucial for the tridiagonal form of the Green's functions
and follows from the fact that
\[  \int_{-\infty} ^{\infty} {d \over dx} \left( x^n P(x) \right) dx =0~~
{\rm if}~~
 \lim_{|x| \rightarrow \infty} x^n P[x] =0~. \]
Thus in particular in the absence of external sources
$\langle {\hat x} (t') x(t)  \rangle =  \langle {\hat x} (t'){\hat
x}(t') \rangle =      0$.
The meaning of the {\em hatted} operators is understood in terms of
the response of the system to an external source.  If one changes the
Hamiltonian by 
\begin{equation}
\delta H = \int dr' {\hat x}(r't) \delta f(r't)~, 
\end{equation}
one then finds \cite{ref:MSR} that the response of any observable $A$
is given by:
\begin{eqnarray}
{\delta \langle A(t) \rangle \over \delta f(t')} |_{\delta f =0}&& =
\langle T \left( A(t) {\hat x}(t') \right) \rangle \nonumber \\
{\delta \langle A(t) \rangle \over \delta f(t') \delta f(t'')} |_{\delta f =0} &&=
\langle T \left( A(t) {\hat x}(t'){\hat x}(t'' \right) \rangle~,
\end{eqnarray}
where $T$ corresponds to the usual time ordered product  operation. 
Since $x(t)$ and $x(t')$ always commute, it is clear that the
  generating functional for both the fluctuation and response
  functions is given by 
\begin{eqnarray}
W[\eta,{\hat \eta}]&& = \ln Z[\eta,{\hat \eta}]= \ln \langle
  T(S[\eta,{\hat \eta}] ) \rangle  \nonumber \\
S[\eta,{\hat \eta}]&& = \exp  \int dt~ dr \left[ x(r,t) \eta(r,t) + {\hat
  x}(r,t){\hat \eta}(r,t) \right]~. \label{eq:gener1}
\end{eqnarray}
In particular the one particle functions are
\begin{equation}
{\delta W \over \delta \eta(t) }|_{\eta={\hat \eta}=0} = \langle x(t)
\rangle~, ~~~
{\delta W \over \delta {\hat \eta}(t)}|_{\eta={\hat \eta}=0}=  \langle{\hat x}(t)
\rangle = 0~. 
\end{equation}
The two particle Green's function matrix is tridiagonal
\begin{equation}
G_2(t) = \left( \begin{array}{cc} 
 U(t)  & D_R(t) \\
 D_R(-t) & 0  
\end{array} 
\right) 
\end{equation}
where
$$ G_{xx}(12) =  U(12) = \langle  x(1) x(2)  \rangle- \langle x(1) \rangle \langle x(2) \rangle~, $$
and
$$G_{x \hx}(12) = D_R(12) =  \langle T \left( x(1)  {\hat x}(2)
\right) \rangle= \Theta(t_1-t_2) \langle   x(1)  {\hat x}(2)
 \rangle~.  $$
  One next doubles the space as in the CTP formalism by introducing
the field  $\Phi(r,t)$ 
\begin{equation}
  \Phi(r,t) = \left( \begin{array}{c} 
 x(r,t) \\ {\hat x}(r,t)  
\end{array}~. 
\right) 
\end{equation}
Then the commutators of $\Phi$ satisfy in the( $ 2 \times 2  $) 
space
\begin{equation}
 \left[ \Phi_\alpha(r,t) , \Phi_\beta(r',t)  \right] = i
 (\sigma_2)_{\alpha \beta} \delta(r-r')~.
\end{equation}
In this larger space
 the equations of motion are
\begin{equation}
{\dot \Phi} = [ \Phi,H]~,
\end{equation}
where the ``non-Hermitian'' operator  $H$ has the form
\begin{equation}
H = \int \gamma^1_a   \Phi_a (1) d1 + {1 \over 2}  \gamma^2_{ab} \int d1~
\Phi_a(1)  \Phi_b(1) + {1 \over 3!} \gamma^3 _{abc} \int d1 ~ \Phi_a
\Phi_b \Phi_c~,
\end{equation}
and the equations of motion in the presence of an external source
$J $ is  
\begin{equation}
-i \sigma_2 {\dot \Phi_a }= J_a(1) +  \gamma^1_a   +   \gamma^2_{ab} \
  \Phi_b(1) + {1 \over 2 } \gamma^3 _{abc} 
\Phi_b \Phi_c~,
\end{equation}
where 
\begin{equation}
J(r,t) = \left( \begin{array}{c} 
 \eta(r,t) \\ {\hat \eta}(r,t)  
\end{array} 
\right)~. 
\end{equation}
The generating functional $W[\eta, {\hat \eta}]$ of Eq. \ref{eq:gener1} is now written in compact
form by writing
\begin{equation}
 S = \exp \left[ \int dr dt \Phi(r,t) J(r,t) \right]~. 
\end{equation}
The expectation value of the equation of motion in the presence of an
external source leads to the  equation
\begin{eqnarray}
-i \sigma_2 {\dot G_1} (1)&& = J(1) +  \gamma^1(1)    +   \gamma^2
  G_1(1) + {1 \over 2 } \gamma^3 [G_2(11) + G_1(1) G_1(1)]~, \nonumber \\
G_m^{J}&& = { \delta^m W  \over \delta J(1) \delta J(2) \cdots \delta
 J(m)}~.
 \label{eq:sd1}
\end{eqnarray}
All the higher n-point functions are obtained by functional
differentiation.  The complete  S-D  equations are derived in
the original MSR paper \cite{ref:MSR}, and a path integral
representation was derived by Jouvet and Phythian \cite{ref:Jouvet}.
These coupled Green's functions are in a $2N  \times  2N$ matrix space and obey first order in time
equations rather than in the usual $N \times N$ space obeying second order equations.  To see 
how to obtain the second order formalism we will simplify the
discussion by  considering a cubic plus quartic 
anharmonic oscillator. 

\sss
{\noindent\bf III. SECOND ORDER MSR FORMALISM}

For the  anharmonic oscillator with equation of motion
\[ {\ddot x} + m^2 x + g x^2 + \lambda x^3 = 0~,  \]
  the above discussion leads
to 
\begin{equation}
H = {\hat x} p - {\hat p} (m^2 x + gx^2 + \lambda x^3)~. 
\end{equation}
This yields  the first order equations 
\begin{eqnarray}
{dx \over dt} && =[x,H] =p~, ~~~
{dp \over dt}  =[p,H] = - (m^2 x + g x^2 + \lambda x^3)~,   \nonumber \\
{d \hx  \over dt} && =[\hx,H] =  \hp (m^2 + 2g x+ 3 \lambda x^2)~,~~~
{d \hp \over dt} = [\hp,H] = -\hx~.
\end{eqnarray}

The first order  equations for the Green's functions are obtained by 
taking the expectation value of 
these equations with sources added and  then using functional
differentiation.  The expectation value is over an initial
distribution function $ \rho[x_0,p_0]$.  A simple example is
is to choose our ensemble of initial conditions at $t=0$  from a
thermal distribution of the free  Hamiltonian, i.e.
\begin{equation}
\rho = N e^{-\beta H_0}~, ~~~ H_0 = {1 \over 2} p^2  + {1 \over 2} m^2 x^2~,~~~ \int \rho dx
dp = 1~. 
\end{equation}
One can show that out of the twelve  nonzero  Green's Functions
in the first order formalism  only two
are independent. These can be chosen as  $G_{xx}$, and $G_{x \hp}$. 

The harmonic oscillator values of  these Green's functions are
easily determined  from the operator solutions
\begin{equation}
x(t) = x_0 \cos mt+ {p_0 \over m} \sin m t~, ~~~ 
\hp(t) = \hp_0 \cos mt - {\hx \over m} \sin mt~,
\end{equation}
and the initial density matrix. 
For the thermal initial conditions described above  we obtain 
\begin{equation}
\langle x(t) x(t') \rangle  = {1 \over \beta m^2} \cos m (t-t')
~,~~~
\langle x(t) \hp(t') \rangle = {1 \over m} \sin m (t-t')~.
\end{equation}
The response function is
\begin{equation}
\langle T(  x(t) \hp(t'))  \rangle = D^0_{ret}(t-t') =\theta(t-t')  {1
\over m} \sin m (t-t')~.
\end{equation}
In general, in the absence of sources 
\[ G_{\hp \hp}(t-t') =0~, ~~~ 
G_{\hp x} (t-t') = G_{x \hp} (t'-t) = D^0_{adv} (t-t')~.  \]
The independent second order equations  (adding sources)  are
\begin{eqnarray}
&&[ {d^2  \over dt^2} + m^2 ] x +g x^2+\lambda x^3  =  j_x~, \nonumber \\
&&[ {d^2  \over dt^2} + m^2 ] \hp  + 2 g x \hp+ 3 \lambda x^2 \hp =  j_{\hp}~.
\end{eqnarray}
Here  $ \hp(0)$ is treated  as an operator -${\delta
\over \delta p}$ when one averages over the initial probability
function in phase space.  

These equations are derivable from the  Lagrangian
\begin{equation}
L_{MSR} = {1 \over 2} \left( \hp ~~[ {d^2  \over dt^2} + m^2 ]~~ x + x~~ [ {d^2
\over dt^2} + m^2 ]~~ \hp \right) + g x^2 \hp+ \lambda x^3 \hp  - j_{\hp}  x - j_x \hp~.
\end{equation} 
The vertices of the classical theory are

\[ \gamma_{\hp x x} =  \gamma_{ x \hp x}= \gamma_{ x x \hp}= 2g~, 
~~~ \gamma_{\hp x x x } =  \gamma_{ x \hp x x}= \gamma_{ x x \hp
x}= \gamma_{ x x x \hp} = 6\lambda ~. \] 

A formal  path integral formalism can be generated for  the generating
functional
\begin{equation}
Z[ j_{\hp}, j_x ] = \langle  \int dx d \hp
e^{-\int L(x,\hp) dt}  \rangle~,
\end{equation}
where the expectation value is over the initial density matrix. 

We  want to compare $L_{MSR}$  with the Lagrangian for the CTP formalism.
The result of the CTP formalism \cite{ref:CTP} is that the action is the
difference of two terms, one for each branch of the closed time path
contour. Explicitly for the anharmonic  oscillator 
  
\begin{eqnarray}
L_{CTP} && = {1 \over 2} \left( x_+  [ {d^2  \over dt^2} + m^2 ] x_+
-  x_{-} [ {d^2
\over dt^2} + m^2 ] x_{-}   \right) \nonumber \\
&& +{1 \over 3} g [ x_+^3 - 
x_{-}^3 ] + {1 \over 4} \lambda [x_+^4 -x_{-}^4]
  -
j_+  x_+ + j_{-}
x_{-}~.
\end{eqnarray} 
Introducing the change of  variables
\begin{equation}
x_+ = x+ {\hbar \hp \over 2}~,~~~ x_{-}  = x -  {\hbar \hp \over 2}~,
\end{equation}
we obtain in this new basis
 
\begin{eqnarray}
\frac{L^{(2)}_{CTP}}{\hbar} && = 
{1 \over 2} \left( \hp ~~[ {d^2  \over dt^2} + m^2 ]~~ x + x~~ [ {d^2
\over dt^2} + m^2 ]~~ \hp \right) + g x^2 \hp + \lambda x^3 \hp
\nonumber \\
&& + {g \hbar^2 \over 12} \hp^3  +{\lambda \hbar^2 \over 4} x \hp^3 
- j_{\hp}  x - j_x \hp~,
\end{eqnarray}
where 
\[  j_{\hp} = j_+-j_{-}~,~~~ j_x = { j_{+} +j_{-} \over 2}~.  \]
This formal manipulation can be justified by first obtaining the SD
equations
 in the $(+,-)$ basis and making a rotation by $\pi/4$ as
discussed in Ref. [1].

We notice that there are five  extra vertices in the quantum case 
\[ \gamma_{\hp \hp \hp}= 
  {g \hbar^2 \over 2}~, ~~~ \gamma_{x \hp \hp \hp}=\gamma_{ \hp x \hp \hp}=
\gamma_{ \hp \hp x \hp} =  \gamma_{ \hp \hp \hp x} ={3\lambda \hbar^2 \over 2}~.
  \]
 not present in the MSR Lagrangian.  
Note that $\hbar$ arises from the fact that the  classical  
$\hp$ is the derivative operator -$d/dp$ which for quantum mechanics
becomes  $
{\hbar \over i}  d/dp $. It may be noted that apart from these extra
vertices in the quantum case, $\hbar$ dependence also enters in the
initial conditions on the Green's functions. This dependence is 
explicit in the commutator contributions and also occers in the $\hbar$
dependence on the initial distribution which is now a Bose-Einstein
distribution rather than a Maxwell-Boltzmann one.  As an illustration consider
the quantum version of Eq. (26) for the Harmonic oscillator with an initial
Bose distribution. Then we obtain
\begin{equation}
x^Q(t) = x_0^Q \cos (mt) + {p_0^Q \over m} \sin (mt).
\end{equation}
From this we find
\begin{eqnarray}
\langle x(t) x(t') \rangle&& = \langle x_0^2 \rangle \cos (mt) \cos (mt') +\langle {p_0^2 \over m^2}  \rangle \sin (mt) \sin (mt') \nonumber \\
&& - i {\hbar \over 2} \sin [m (t-t')],
\end{eqnarray}
where
\begin{equation}
\langle x_0^2 \rangle = {\hbar \over 2 m} + { \hbar \over m} [\exp(\hbar \beta m)-1]^{-1}.
\end{equation}
and
\[ \langle p_0^2 \rangle = m^2 \langle x_0^2 \rangle. \]

So we see that $\hbar$ enters in many ways in the initial conditions
for the Green's functions, and also in the structure of the S-D equations
where certain vertices are proportional to $\hbar^2$. 
The  way one derives the S-D equations from the action is identical for
both classical and quantum mechanics. Thus we obtain the same
structure classically, but there are fewer vertex functions. 

We now
derive the S-D equations for a generic cubic self-interacting field
theory whether classical or quantum.
 For 
$N$ interacting scalar fields, we
introduce the column vector $\Phi_\alpha$ composed of $\Phi^1_i= \phi_i$ and
$\Phi^2_i = {\hat \pi}_i$ where $i=1,2, \cdots N$. We also need the
metric $g_{\alpha \beta}$ which is just $\sigma^1_{\alpha \beta}$ as far as
connecting the $\Phi^1$ and $\Phi^2$ sectors.  Then we can write
generically for cubic interactions:
\begin{equation}
 {\cal L} = {1 \over 2 } \phi^\alpha (D_0^{-1})_{\alpha \beta}
 \Phi^\beta + {1 \over 3!} \gamma_{\alpha \beta \rho}  \Phi^\alpha \Phi^\beta
 \Phi^\rho  -
 J_\alpha  \Phi^\alpha~,
\end{equation}
where
\[ D^{-1}_{0~~\alpha \beta}(x-y)  = g_{\alpha \beta}[ \Box+ m^2 ] \delta
(x-y)~. \]  
The generating functional is formally
\begin{equation}
Z = \langle  T(\exp [J_{\alpha} \Phi^\alpha])  \rangle  = e^{W[J]}~.
\end{equation}
Defining the ``classical''  field $\Phi^\alpha$ and the connected 2
point function $G$  via
\[   \Phi^\alpha = {\delta W \over \delta J_\alpha}~,~~ G(1,2)_{\alpha \beta} = {\delta \Phi_\alpha(1) \over \delta
J_{\beta}(2)}~, 
 \]
one has that in the presence of sources
\begin{eqnarray}
 && D^{-1}_{0 ~~\alpha \beta} \Phi^\beta (1) + {1 \over 2} \gamma_{\alpha \beta
  \rho}[{\delta \Phi^\beta(1) \over \delta J_\rho(1)} + \Phi^\beta(1)
  \Phi^\rho(1) ] = J_\alpha(1)~, \nonumber \\
 && D^{-1}_{0 ~~\alpha \beta} G ^{\beta \rho}  (12) + {1 \over 2} 
\gamma_{\alpha \beta
  \sigma }[{\delta G^{\beta \sigma} (11) \over \delta J_\rho(2)} +
  \Phi^\beta(1)
G^{\sigma \rho} (12) +\Phi^\sigma(1)
  G^{ \beta  \rho} (12)  ] \nonumber \\
&& = \delta_\alpha^\rho \delta(1-2)~.  \label{eq:dy1} 
\end{eqnarray}
To obtain the S-D equation for the inverse two point function we first
use the connection between the connected 3-point function and the 1-PI
vertex function \cite{ref:CJT} obtained by first 
 Legendre transforming from $J$ to $ \Phi $ 
 and  using the chain rule
\begin{equation}
{\delta \over \delta J_\alpha (1)} 
= \int d2~  G (12)^{\alpha \beta} {\delta \over \delta \Phi(2)^\beta}~, 
\end{equation}
\begin{eqnarray}
&&\langle T \biggl( \Phi_\alpha(1) \Phi_\beta(2) \Phi_\rho(3) \biggr) \rangle_{c}
={\delta G_{\alpha
 \beta} (12) \over \delta J^\rho(3)} \nonumber \\
&& = \int d4~d5~d6 G_{\alpha \alpha'}(14) G_{\beta \beta'}(25) G_{\rho
\rho'} (3,6) \Gamma^{\alpha' \beta' \rho'} (456)~.
\end{eqnarray}
where the 1-PI three point function is defined by 
 \begin{equation}
 \Gamma_ {\alpha \beta \rho} (123) = {\delta G^{-1}_{\alpha \beta}
 (1,2) \over \delta \Phi^\rho(3)}~.
\end{equation}
The usual S-D  is obtained by multiplying
Eq. \ref{eq:dy1}  on the right by  $ G^{-1}$ to
obtain
\begin{equation}
G^{-1}_{\alpha \beta}(1,2) = D^{-1}_{0 \alpha \beta}(1,2)+ \Sigma_{
\alpha \beta}(1,2)+   \gamma_ {\alpha \beta \rho} \Phi^\rho(1)
\delta(1-2)~,
\end{equation}
and 
\begin{equation} 
\Sigma_{\alpha \beta}(1,2)= {1 \over 2} 
\int d3d4 \gamma_{\alpha \rho \sigma}  G^
{ \rho \lambda} (13)G^{\sigma \nu }(14) \Gamma(342)_{ \lambda \nu \beta}~.
\end{equation}

We also  have
\begin{eqnarray}
\Gamma_{\alpha \sigma \rho} (1,2,3) 
= \gamma_{\alpha \sigma \rho} \delta (1-3) \delta(1-2) + {\delta
\Sigma_{\alpha \sigma}(1,2)
\over \delta \Phi^\rho(3)}~.
\end{eqnarray}
However $\Sigma$ is a proper self energy and  can be related
 \cite{ref:Baym}  to the effective action $\Gamma_2[G]$ for the 2-PI
irreducible graphs of the theory via
\begin{equation}
   \Sigma(12)_{\alpha \beta} = 2  {\delta \Gamma_2[G] \over \delta
   G^{\alpha \beta} (1,2)}~,
\end{equation}
so that  it is just a function of the bare vertices and the full
Green's functions. Using 
\begin{equation}
  {\delta  \Sigma(12)_{\alpha \beta} \over \delta \Phi_\rho(3)}  =
   \int d5 d6   {\delta \Sigma(12)_{\alpha \beta}  \over \delta
   G^{\sigma  \lambda} (5,6)} { \delta
   G^{\sigma  \lambda} (5,6) \over \delta \Phi_\rho(3)}~, 
\end{equation}
we obtain that
\begin{eqnarray}
&&\Gamma_{\alpha \beta \nu} (1,2,3)= \gamma_{\alpha \beta \nu}
\delta (1-2) \delta (1-3)
 \nonumber \\
&& -  \int d4 ~d5~d6~d7~  \Gamma_{\alpha \rho  \sigma }(145) G^{\rho \eta}(46) G^{\sigma
\lambda} (57)H(67;23)_{\eta  \lambda; \beta \nu}~. 
\end{eqnarray}
The scattering Kernel $H$ is defined by 
\[    {\delta \Sigma_{\alpha \beta}(12)  \over \delta
   G^{\sigma  \lambda} (5,6)} = H(12 56)_{\alpha \beta \sigma \lambda}~. \]

Self consistent approximations \cite{ref:Baym, ref:CJT}  are determined by keeping a certain
class of graphs in $\Gamma_2[G]$, the sum of all 2-PI graphs made from
bare vertices and full propagators. For cubic interactions,
 the Bare Vertex Approximation (BVA) is
obtained by keeping the  graph 
\[ \int d1 d2  \gamma_{ijk} G^{il}(12) G^{jm} (12))G^{kn}(12)
\gamma_{lmn}~, \]
which then leads to the self energy being the one loop diagram, and
the scattering Kernel being single particle exchange.  By using the
variational definitions of $\Sigma$ and $H$ one is guaranteed an
internally consistent  approximation.
As an example of the difference between the quantum and classical S-D
equations, let us restrict ourselves to a cubic anharmonic oscillator
and make the bare vertex approximation.  The S-D equation for the correlation
function
 becomes
\begin{equation}
G_{x x} (12) = G^0_{x x} (1 2) + \int d3 d4
G^{(0)}_{xi} (13)
\gamma_{ijk} G^{jl}(34) G^{km}(34) \gamma_{lmn} G_{n x }(42)~,
\end{equation}
which can be identified with Kraichnan's Direct Interaction Approximation
\cite{ref:kra}.
Expanding we can write this in terms of classical and quantum
contributions.
We have in symbolic form
\begin{equation}
G_{xx} = G^0_{xx} + G^0_{xx} \Sigma_A^{cl} G_{\hp x} + G^0_{x \hp}
\Sigma_R^{cl}G_{xx} 
+G^0_{x \hp} [\Sigma_F^{cl} +\Sigma_F^{Q}] G_{\hp x}~. 
\end{equation}
The classical contributions to the self-energy in the BVA are
\begin{eqnarray}
\Sigma_F^{cl} (12)&&  = {1 \over 2} \gamma^2_{\hp xx} G_{xx}^2(12)~, 
~~\Sigma_R^{cl} (12)= 
\gamma^2_{\hp x x} G_{x \hp}(12)  G_{xx}(12)~, \nonumber \\
\Sigma_A^{cl} (12) && =
\gamma^2_{\hp x x} G_{\hp x }(12)  G_{xx}(12)~,
\end{eqnarray}
and the quantum contribution is 
\begin{equation}
\Sigma_F^Q (12)  = {1 \over 2} \gamma_{\hp xx} \gamma_{ \hp \hp \hp} 
 [G_{x \hp}^2(12) +
 G_{ \hp x}^2(12)]~,
\end{equation}
which has formal order of $\hbar^2$ since $\gamma_{ \hp \hp \hp}$ is of
order $\hbar^2$.

We have recently shown that the {\em classical} BVA gives excellent
agreement with Monte Carlo simulations in $1+1$ dimensional $\phi^4$
field theory \cite{ref:new}.

\sss
\sss
{\noindent\bf IV. CONCLUSIONS}

We have shown how  the classical S-D equations can be derived from
the quantum S-D equations by comparing the CTP formalism with the MSR
formalism.  Our hope is that this result  will allow researchers to 
 make clearer the connection
between quantum field thory and classical field theory at high
temperatures.  We are in the process of comparing the BVA for the 
quantum and classical cases at high and low temperature to understand
for what range of temperatures the classical approximation is valid.

\end{document}